\documentclass[conference]{IEEEtran}
\IEEEoverridecommandlockouts
\usepackage{cite}
\usepackage{amsmath,amssymb,amsfonts}
\usepackage{algorithmicx}
\usepackage{graphicx}
\usepackage{textcomp}
\usepackage{xcolor}
\usepackage{multirow}
\usepackage{tikz}
\usepackage{comment}
\usetikzlibrary{positioning, fit}

\def\BibTeX{{\rm B\kern-.05em{\sc i\kern-.025em b}\kern-.08em
    T\kern-.1667em\lower.7ex\hbox{E}\kern-.125emX}}
\begin{document}

\title{Combining Forensics and Privacy\\ Requirements for Digital Images
\thanks{This work has been funded in part by the French National Research Agency (ANR-18-ASTR-0009), ALASKA project: https:// alaska.utt.fr, and by the French ANR DEFALS program (ANR-16- DEFA-0003).}
}

\author{\IEEEauthorblockN{Pauline Puteaux}
\IEEEauthorblockA{\textit{Univ. Montpellier} \\
\textit{CNRS, UMR 5506 LIRMM}\\
Montpellier, France \\
pauline.puteaux@lirmm.fr}
\and
\IEEEauthorblockN{Vincent Itier}
\IEEEauthorblockA{\textit{IMT Lille-Douai, Institut Mines-Télécom,}\\\textit{Centre for Digital Systems, F-59000 Lille, France} \\
\textit{Univ. Lille, CNRS, Centrale Lille, Institut Mines-Télécom,}\\\textit{UMR 9189 CRIStAL, Lille, France}\\
vincent.itier@imt-lille-douai.fr}
\and
\IEEEauthorblockN{Patrick Bas}
\IEEEauthorblockA{\textit{CNRS, Univ. Lille,} \\
\textit{Centrale Lille, UMR 9189 CRIStAL}\\
Lille, France  \\
patrick.bas@centralelille.fr}
}

\maketitle

\begin{abstract}
This paper proposes to study the impact of image selective encryption on both forensics and privacy preserving mechanisms. The proposed selective encryption scheme works independently on each bitplane by encrypting the $s$ most significant bits of each pixel. We show that this mechanism can be used to increase privacy by mitigating image recognition tasks. In order to guarantee a trade-off between forensics analysis and privacy, the signal of interest used for forensics purposes is extracted from the $8-s$ least significant bits of the protected image. We show on the CASIA2 database that good tampering detection capabilities can be achieved for $s \in \{3,\dots,5\}$ with an accuracy above 80\% using SRMQ1 features, while preventing class recognition tasks using CNN with an accuracy smaller than 50\%.   
\end{abstract}

\begin{IEEEkeywords}
Forensics, Privacy, Visual confidentiality, Selective encryption, Trade-off.
\end{IEEEkeywords}

\section{Introduction}
Image exchanges represent a large amount of Internet usage nowadays. This trend goes hand in hand with privacy requirements since the transmission can be spied on public channels. Therefore, it has been proposed to encrypt these images in order to hide their content, making them visually confidential to unauthorized users. Some encryption methods have been specifically designed for images in order to preserve their format and their size and allowing their visualization after encryption. Allowing visualization is interesting to let users being able to see that an image is present, but its access is restricted. Moreover, selective encryption, which only encrypts a fraction of image information, allows us to visualize a level of details of the image as a function of the encrypted information~\cite{van2002techniques}. In addition, visualization may be authorized only on a certain part of the image. Encryption can be then done partially, for example only on human faces, for privacy concerns. In this context, partial encryption can be selective~\cite{rodrigues2006selective}. Nevertheless, for end users such as cloud platforms or image based social networks, encrypted images are not convenient to work with. Indeed, using a classical encryption scheme, the targeted platform is not able to decide whether an image respects the terms of usage or not. In particular, it cannot check its integrity as this is done in the clear domain~\cite{christlein2012evaluation}. In order to preserve privacy while enabling analysis in the encrypted domain, homomorphic encryption has been proposed. This approach can be used for SIFT detection for example~\cite{hsu2011homomorphic}. However, homomorphic encryption schemes are computationally intensive, which avoids complex operations from being carried out, and requires more storage. %(usually two times the image size in memory). % PB: I'm not sure of this last statement. 
On the contrary selective encryption is fast and does not expand the original image size. With such an approach, a part of the image content is encrypted, while the other one remains in clear, {\em i.e.} non-encrypted, and can be then analyzed. This could introduce a security breach and image content privacy is thus questionable.

\begin{figure}[!t]
\begin{center}
\resizebox{\columnwidth}{!}{
\input{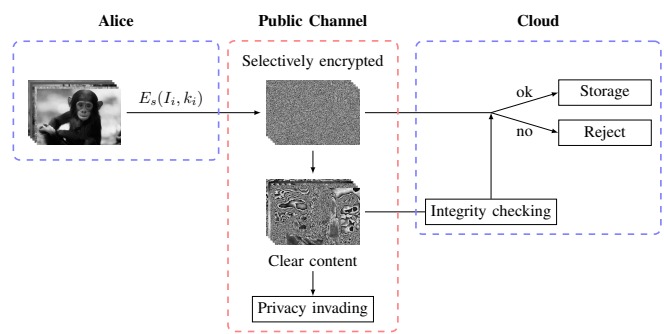}
}
\end{center}
\vspace{-2mm}
\caption{Trade-off between privacy preservation and integrity check in the context of selectively encrypted image exchanges throughout a public server.}
\label{fig:context}
\end{figure}

In this paper, we study how it is possible to use the framework of selective encryption in order to reach a trade-off between privacy preservation and integrity check. %in the context of selectively encrypted image exchanges throughout a public server, of which 
An illustration of an application scenario on a public server is depicted in Fig.~\ref{fig:context}. From original images, several bit-planes are encrypted, from most significant to least significant bits. A forensics analysis based on the extraction of SRMQ1 features is then conducted to detect if a selectively encrypted image has been tampered or not. In addition, a privacy evaluation is carried out in order to assess the visual confidentiality of a selectively encrypted image. This is done in terms of recognizability by predicting the image class. 

%The main drawback is that the clear part, by definition and according to the Kerckhoffs's principle, is available. This trade-off is illustrated in Fig.~\ref{fig:context}.

% parler d'integrity check vs visual confidentiality

The rest of this paper is organized as follows. Section~\ref{sec:method} describes our proposed approach to analyze the influence of selective encryption on both image forensics and privacy preserving mechanisms. Experimental results are presented in Section~\ref{sec:results}. Finally, the conclusion is drawn in Section~\ref{sec:ccl}.

\section{Proposed approach\label{sec:method}}
In this section, we describe our proposed approach to analyze the trade-off between privacy and tampering detection in the context of selective encryption. Selective encryption consists to encrypt the most significant bit-planes (MSB) of an image, while keeping the least significant bit-planes (LSB) in clear. In order to perform forensics on selectively encrypted images, we focus on the residual information of the image.% for classification as authentic or tampered. 
Moreover, for privacy evaluation, we are interested in assessing the recognizability of a selectively encrypted image by automatically predicting the class of the content. Note that our approach is detailed for an application to gray level images, but can be easily extended to RGB color images.  

\subsection{Selective encryption}
Let us consider a gray level input image of $m\times n$ pixels. Each pixel $p(i,j)$ from this image, $0\leq i < m$ and $0\leq j < n$, is made of 8 bits and defined as:
\begin{equation}
p(i,j)= \sum_{k=0}^{7}p^k(i,j)\times 2^{7-k},
\end{equation}
where $p^k(i,j)$ is the bit of index $k$.

One can note that the smaller the index $k$, the more significant the associated bit. For privacy requirements, the input image is encrypted in order to ensure the visual security of its content. Moreover, depending on the application, it should be interesting to be able to preserve a part of the image in clear. In this context, encryption is selectively performed. Only a fixed number $s$ of bit-planes are encrypted and the remaining $8-s$ ones are kept in clear. Encryption is then performed from the most ($k=0$) to the least ($k = s-1$) significant bit-plane to encrypt (from MSB to LSB). An encryption key is used as a seed for a cryptographically secure pseudo-random number generator to obtain a pseudo-random sequence of $s\times m\times n$ bits $b^k(i,j)$, with $0\leq k < s$. For each bit-plane to encrypt, each bit $p^k(i,j)$ is XOR-ed with the associated bit in the pseudo-random sequence to generate an encrypted bit $p^k_E(i,j)$:
\begin{equation}
p^k_E(i,j) = p^k(i,j) \oplus b^k(i,j).
\end{equation}

\subsection{Tampering detection using residuals}
Tampering detection aims to decide whether or not an image has been altered by local modification. Most common forgeries are cloning (copy/move from a single image) and splicing (copy/paste between several images). If typical image forensics techniques use as inputs the whole image to be analyzed, this strategy is not the best for encrypted images since the encryption adds a noise of strong magnitude. This noise could also alter the extraction of significant features for a classification as authentic or tampered. 

In order to perform a forensics analysis from selectively encrypted images, all the encrypted bit-planes, of index \mbox{$0\leq k < s$}, should be discarded in a pre-processing step. According to the Kerckhoffs' principle, we can assume that the number $s$ of encrypted bit-planes is known.  Therefore, for each pixel $p_E(i,j)$, the encrypted bits are set to zero using bitwise shift operations to obtain a value $p_0(i,j)$:
\begin{equation}\label{eqn:zeros}
p_0(i,j) = (p_E(i,j) \ll s) \gg s.
\end{equation}

In doing so, only the non-encrypted least significant bit-planes are considered. One can note that these bit-planes should be the most relevant for the classification task because they are directly linked to the image residuals. Steganalysis domain falls within the search of weak signals in image residuals. Due to the intrinsic properties of traces left by image forgery, steganalysis approaches can be applied to image forensics~\cite{chierchia2014bayesian,cozzolino2014image}.

In this context, one of the most popular feature extractor is the Spatial Rich Model (SRM)~\cite{fridrich2012rich}. Because it uses the statistics of neighboring noise residuals, it is widely employed for steganalysis, but can be also used for tampering detection. Indeed, noise residuals correspond to high frequency components of an image. They capture the dependency changes due to the tampering operation, in both horizontal and vertical directions. The SRM begins by the computation of the residuals. During this step, the input image is filtered by several high-pass filters to generate residual images with different shapes and orientations. After that, a quantization and a truncation steps are performed. Finally, an output feature vector with 37,561~residuals is obtained, whatever the size of the input image. The main drawback of SRM is that it leads to a high computational complexity. Therefore, in order to deal with this issue, a simplified version called SRMQ1 can be used instead. With this feature extractor, the output feature vector only contains 12,753~residuals. 

For classification, an implementation of ridge regression using Least Square Minimum-Residual (LSMR) optimization method is used, due to its
low computational complexity and low memory requirements~\cite{fong2011lsmr,cogranne2015ensemble}. Two classes are considered for classification: authentic, {\em i.e.} with no falsification, and tampered, when there are forgeries due to cloning or splicing operations.

\begin{figure*}[ht]
\centering
\includegraphics[width=\linewidth]{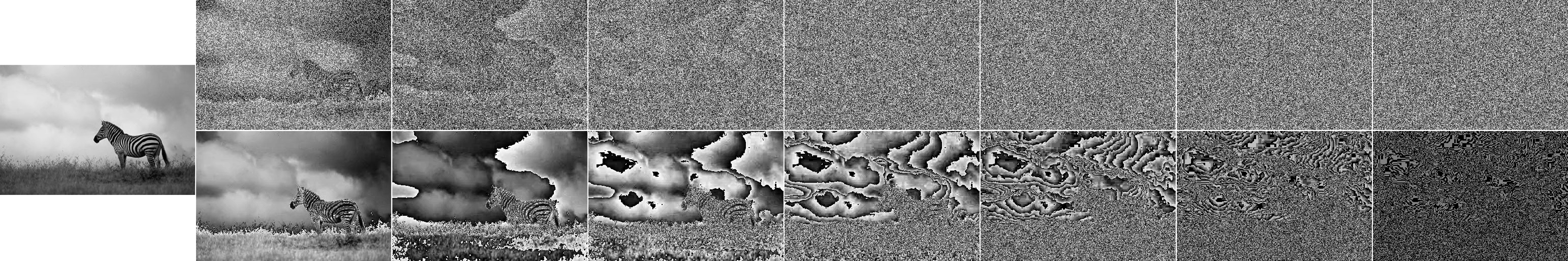}
\vspace{-2mm}
\caption{Illustration using the luminance component of the original image {\em Au\_ani\_00001} from the CASIA2 database~\cite{dong2013casia}: first row)~Selectively encrypted images depending on the number $s$ of encrypted bit-planes, from MSB to LSB and with $1\leq s\leq 7$; second row)~Images obtained by setting to zero the $s$~encrypted bit-planes of associated selectively encrypted images (images were standardized for the recognizability task; this also allows a better visualization of the significant information for classification).\label{fig:ex}}
\end{figure*}

\subsection{Privacy evaluation of selectively encrypted images\label{subsec:privacy}}
The selective encryption allows us to hide some levels on details of the image. The proposed tampering detection relies on the LSB that are not encrypted. Using a non full encryption method could lead to privacy leak using the clear content of the image. Therefore, we aim to know what is the trade-off between the visual confidentiality, which assesses the privacy, and the tampering detection.

The assessment of the visual confidentiality of an image is a difficult task. Indeed, it is known that usual quality metrics, such as PSNR or SSIM, are not relevant for assessing a perceptual low quality. A low score does not point out if the image has just a low quality or if the content can not be recognized \textit{i.e.} if the encryption preserves the privacy. Some perceptual metrics based on subjective evaluations were proposed. In the context of privacy evaluation, the main drawback of these metrics is that they focus on perceptual quality. Recently, Hofbauer~{\it et al.}~\cite{HOFBAUER2021128} have proposed an encrypted image database with subjective recognition ground truth and analyzed the correlation of subjective scores with some state of the art metrics. They conclude that the evaluation of image quality and the evaluation of the content recognizability are two really different tasks and, therefore, the visual quality metrics should not be used to assess content recognizability.

We propose to evaluate the privacy of a selectively encrypted image by trying to automatically predict its class. Our assumption is that if an algorithm can automatically predict the content of an image, then the encrypted image is leaking some visual information\textit{ i.e.}. %low frequency features can be recognizable. 
Image content classification methods have achieved high performance thanks to convolutional neural networks (CNN). Therefore, we propose to assess the recognizability of selectively encrypted image by training a model which predicts the class of the image. 

As for forensics analysis, the $s$ encrypted bitplanes should be set to zero using Eq.~\eqref{eqn:zeros}. This boils down to work directly on high frequencies and short intensity range. In order to exploit the low dynamic, images are standardized before to be passed as input to the model.

Finally, to assess the classification score we use the accuracy score which measures how much data have been well predicted. As visual confidentiality is inherently linked to recognizability, we propose to define a privacy index such as $1 - accuracy_\text{\em~recognizability}$. Indeed, the more easily the content of an image is recognizable, the lower the level of visual confidentiality: recognizability and privacy are consequently antagonist.

\begin{comment}
\renewcommand{\arraystretch}{1.25}
\begin{table*}[ht]
\centering
\caption{\label{tab:accuracy1} Accuracy for tampering detection using SRM and SRMQ1 as a function of the number of encrypted bitplanes (from MSB to LSB).}
\begin{tabular}{|c|c|c|c|c|c|c|c|c|c|}
\hline
\multirow{2}{*}{\begin{tabular}[c]{@{}c@{}}Feature\\ extractor\end{tabular}} & \multicolumn{9}{c|}{Number $s$ of encrypted bitplanes}           \\ \cline{2-10} 
                                                                             & 0    & 1    & 2    & 3    & 4    & 5    & 6    & 7    & 8    \\ \hline
\multicolumn{10}{|c|}{Without pre-processing}\\ \hline
SRM~\cite{fridrich2012rich}  &   0.91   &   0.80   &   0.75   &  0.68    &            0.65 &   0.62   &   0.59  &   0.54 & 0.50  \\ \hline
SRMQ1~\cite{fridrich2012rich} & 0.90 & 0.81 & 0.76 & 0.69  & 0.62 & 0.61 & 0.55 & 0.54 & 0.50 \\ \hline
\multicolumn{10}{|c|}{With encrypted bitplanes set to zero}\\ \hline
SRM~\cite{fridrich2012rich} &   0.91   &   0.89   &   0.88   &  0.86    &         0.85   &   0.80   &   0.78   &   0.72 & 0.50   \\ \hline
SRMQ1~\cite{fridrich2012rich} & 0.90 & 0.87 & 0.87 & 0.86 & 0.84 & 0.81 & 0.79 & 0.72 & 0.50 \\ \hline
\end{tabular}
\vspace{-2mm}
\end{table*}
\end{comment}

\renewcommand{\arraystretch}{1.25}
\begin{table*}[ht]
\centering
\caption{\label{tab:accuracy1} Accuracy for tampering detection using SRMQ1~\cite{fridrich2012rich} as a function of the number of encrypted bitplanes (from MSB to LSB).}
\begin{tabular}{|c|c|c|c|c|c|c|c|c|c|}
\hline
\multirow{2}{*}{\begin{tabular}[c]{@{}c@{}}Feature\\ extraction\end{tabular}} & \multicolumn{9}{c|}{Number $s$ of encrypted bitplanes}           \\ \cline{2-10} 
                                                                             & 0    & 1    & 2    & 3    & 4    & 5    & 6    & 7    & 8    \\ \hline
                                                                             Without pre-processing & 0.90 & 0.81 & 0.76 & 0.69  & 0.62 & 0.61 & 0.55 & 0.54 & 0.50 \\ \hline
                                                                            With encrypted bitplanes set to zero & 0.90 & 0.87 & 0.87 & 0.86 & 0.84 & 0.81 & 0.79 & 0.72 & 0.50 \\ \hline
\end{tabular}
\vspace{-2mm}
\end{table*}

\renewcommand{\arraystretch}{1.25}

\begin{table*}[ht]
\centering
\caption{\label{tab:accuracy} Accuracy for the recognizability task as a function of the number of encrypted bitplanes (from MSB to LSB).}
\begin{tabular}{|c|c|c|c|c|c|c|c|c|c|}
\hline
\multirow{2}{*}{\begin{tabular}[c]{@{}c@{}}Image\\ database\end{tabular}} & \multicolumn{9}{c|}{Number $s$ of encrypted bitplanes}           \\ \cline{2-10} 
                                                                             & 0    & 1    & 2    & 3    & 4    & 5    & 6    & 7    & 8    \\ \hline
CASIA2~\cite{dong2013casia}  & 0.76 & 0.68 & 0.56 & 0.45 & 0.36 & 0.29 & 0.25 & 0.26  & 0.14 \\ \hline
Intel~\cite{intel}  & 0.93 & 0.89 & 0.82 & 0.73 & 0.60 & 0.52 & 0.45 & 0.49  & 0.17 \\ \hline
Cifar10~\cite{krizhevsky2009learning} & 0.87 & 0.68 & 0.52 & 0.37 & 0.23 & 0.13 & 0.09 & 0.10  & 0.10 \\ \hline
\end{tabular}
\vspace{-2mm}
\end{table*}

\section{Experimental results\label{sec:results}}
In this section, we present experimental results assessing the feasibility of combining forensics and privacy requirements for digital images. First, we provide an illustration of selectively encrypted images and standardized images in order to visualize the high frequency information. We then describe the training and the classification results obtained for the tampering detection and the recognizability tasks considering selectively encrypted images. Finally, we discuss the trade-off between tampering detection and privacy.

\subsection{Examples of selectively encrypted images}
In Fig.~\ref{fig:ex}, we first present the luminance component of the original image {\em Au\_ani\_00001} from the CASIA2 database~\cite{dong2013casia}. As an illustration, in the first row of the figure, we display selectively encrypted images obtained by encrypting $s$ bit-planes of this image, from MSB to LSB and with $1\leq s\leq 7$ (from left to right). One can notice that as soon as at least two bit-planes are encrypted, it is visually difficult to recognize the original image content. Indeed, in this example, distinguishing the silhouette of the zebra is not an easy task. In the second row, the presented images have been obtained by setting to zero the $s$~encrypted bit-planes of the selectively encrypted images and by performing a classical image standardization. Even after this process, the content of the original image is unrecognizable when at least five bit-planes are encrypted. This kind of images are taken as input of the CNN for the recognizability task. Moreover, they illustrate the significant information for classification in both forensics analysis and privacy evaluation tasks.

\subsection{Forensics analysis}
The CASIA2 database~\cite{dong2013casia} consists of authentic and tampered images (cloned or spliced) on JPEG or TIFF formats and with a size between $240\times 160$ and $900\times 600$~pixels~\cite{dong2013casia}. One can note that: 1)~tampered images have been generated using a subset of authentic images, and 2)~several tampered images have been issued from the same authentic images. In order to remove this bias in the construction of the database, we have randomly picked 1,000~authentic images and 1,000~tampered images in the full database making sure that there is no overlap between images, {\em i.e.} an image content only appears one time. Then, we have designed eight associated databases of selectively encrypted images, by encrypting between 1~to~8~bit-planes from MSB to LSB. After that, each of them has been processed separately. Into each database, images have been split into two balanced subsets with as many authentic images as tampered images: 80\%~of the images have been used for training and the remaining 20\%~for test. As feature extractor, we have used SRMQ1~\cite{fridrich2012rich}. In Table~\ref{tab:accuracy1}, we present the accuracy scores obtained during the test phase as a function of the number of encrypted bit-planes. First of all, on clear images ({\em i.e.} without encryption), we can see that the accuracy is equal to 0.90 even using a feature extractor as simple as SRMQ1. 
To put this result in perspective with the state of the art, one of the best performing method~\cite{rao2016deep} uses CNN to achieve 0.97 accuracy, using 1:6 train to test ratio. Moreover, we can see that the pre-processing step consisting in discarding the encrypted bit-planes is relevant. If the performances are still quite good for $s=1$ and $s=2$ without pre-processing, the accuracy score falls significantly as soon as three bit-planes are encrypted. With encrypted bit-planes set to zero, what is particularly interesting is that, even with a reduced number of bit-planes in clear, accuracy remains high. Indeed, it is higher than 0.80 considering at least three bit-planes in clear and remains higher than 0.70 with only one or two bit-planes in clear. Therefore, with a very small amount of information on high frequencies, the tampering detection task is possible. 
Note that the results obtained using SRMQ1 are comparable with those achieved using SRM, which highlights that the simplified version of SRM can be used in practice.

\subsection{Recognizability}

The recognizability of image content is assessed by automatically predicting the image class.

The CASIA2 database also provides coarse categories for image content: animals, architecture, art, character, indoor, nature, plants, text and sec. We choose to use the $7,491$~authentic images of the CASIA2 database for this task because authentic images are well labeled and do not contain falsification on which the model may focus. The number of images is relatively small thus, we propose to use the VGG11~\cite{SimonyanZ14a} network pre-trained on ImageNet~\cite{imagenet_cvpr09} as our baseline model. The database is randomly split into two subsets with 80:20 ratio for train and test. Images are cropped at their center to a size of $224 \times 224$ pixels to be passed as input of the model. The model is fine tuned using the train set, it converges quickly and it is stopped before overfitting. The model can predict CASIA2 classes with an accuracy of $0.76$ on the test set. This task is difficult because classes are not well defined and there are some overlap. Nevertheless, it shows that the model is able to predict CASIA2 classes on clear images.

In order to see if the content is still recognizable after the encryption of the $s$ most significant bitplanes, the baseline model is fine tuned using the same training set in which images are selectively encrypted. As we have to consider that the number $s$ is known, the best case for image classification is to work directly on the clear bits of the image. Therefore, image pixels are transformed using Eq.~\eqref{eqn:zeros}. In practice, as we want to standardize the model inputs, it is sufficient to apply the left shift operation $p_E(i,j) \ll s$ and then standardize images using classical image standardization.  We also perform these results on the selectively encrypted dataset. With $s=1$, the accuracy of the recognizability task is only of $0.37$, and for $s>1$, the accuracy is close to $0.14$. Indeed, the model does not directly converge toward the extraction of features that do not rely on the $s$ encrypted bits. Thus, it tends to classify all images into the most common class, \textit{i.e.} the ``animal" class which represents 14\% of the base. The fine tuning and testing phases have been independently done for $s \in \{0,7\}$, where $s=0$ means the image is in clear. The obtained results are reported in Table~\ref{tab:accuracy}. We also present results we have obtained using the Intel image classification~\cite{intel} and the Cifar10~\cite{krizhevsky2009learning} databases which were designed for image classification. The total images in each class is balanced. Intel image classification database contains $17,034$ images of  $150 \times 150$ pixels ($14,0034$ for train and $3,000$ for test) separated into $5$ classes: ``sea", ``mountain", ``buildings", ``forest", ``street" and ``glacier". CIFAR10 database is composed of $6,000$ images of $32 \times 32$ pixels ($5,000$ for train and $1,000$ for test) belonging to one of $10$ classes: ``airplane", ``automobile", ``bird", ``cat", ``deer", ``dog", ``frog", ``horse", ``ship" and ``truck". The recognizability task performs better on the Intel database because its classes are well separated, whereas in CIFAR10 there are classes that are close such as ``birds" and ``plane" or ``automobile" and ``truck". Note that the trend observed on the CASIA2 database is firmly established.

\subsection{Trade-off between tampering detection and privacy}
\begin{figure}[ht]
\centering
\includegraphics[width=0.95\columnwidth]{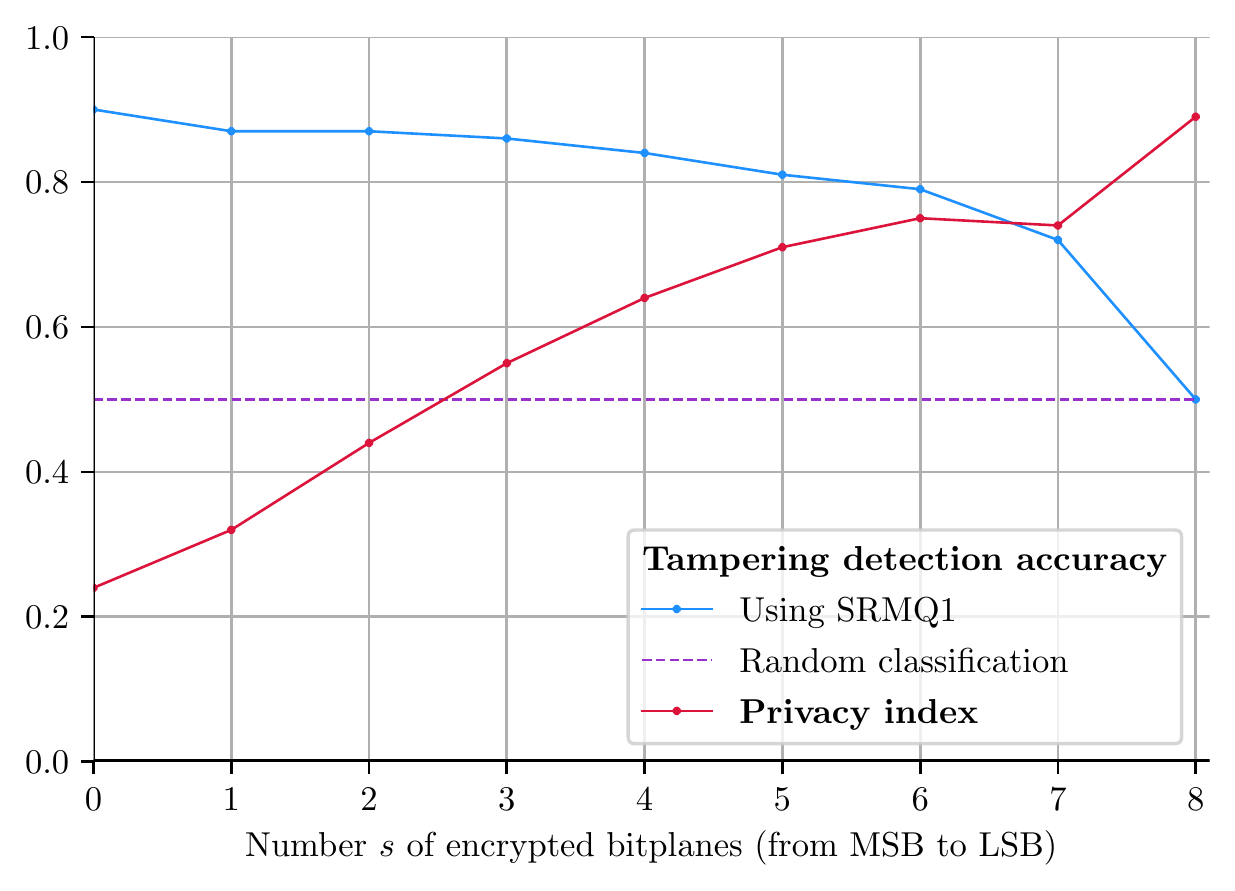}
\vspace{-2mm}
\caption{Trade-off between tampering detection accuracy and privacy as a function of the number $s$ of encrypted bitplanes (from MSB to LSB).\label{fig:tradeoff}}
\end{figure}

In Fig.~\ref{fig:tradeoff}, we illustrate the trade-off between tampering detection accuracy and privacy, as a function of the number $s$ of encrypted bit-planes. These results were obtained using images from the CASIA2 database. On the one hand, we can see that, from one to five encrypted bit-planes, the tampering detection accuracy is very good (higher than 0.8). On the other hand, the privacy index (computed from recognizability accuracy, as explained in Section~\ref{subsec:privacy}), is higher than 0.5 as long as at least three bit-planes are encrypted. This means that the classification rate is low for recognizability, {\em i.e.} the class of the image is mis-predicted on average. Therefore, this highlights that an interesting trade-off for combining tampering detection and privacy is achieved for three to five encrypted bit-planes. In particular, when five bit-planes are encrypted, tampering detection accuracy is equal to 0.81 and the privacy index is equal to 0.71. Fig.~\ref{fig:ex} illustrates the fact that it is very difficult to visually recognize the content of the selectively encrypted image when five bit-planes are encrypted, even by considering the associated standardized image. Moreover, depending on the application, it can be interesting to favor one of the other classification task (integrity check {\em vs} visual confidentiality).

\vspace{-2mm}

\section{Conclusion\label{sec:ccl}}
In this paper, we have performed an analysis on selectively encrypted images to observe the trade-off between tampering detection and privacy. We have shown that SRMQ1 features can be used for a forensics analysis of selectively encrypted images. Moreover, privacy has been assessed experimentally by measuring the recognizability of an image content after encryption using a CNN. According to our experiments, an accuracy of more than $80\%$ for tampering detection is achieved when $s=0$ to $s=5$ bitplanes are encrypted, whereas the visual confidentiality is ensured as soon as $s=3$ bitplanes are encrypted. 

In future work, we are interested by improving the classification performances during the forensics analysis using more specific tools, as those used for non-encrypted images. These approaches often rely on deep learning. Therefore, they may require using a larger database than CASIA2. It also could be interesting to investigate the tampered areas localization too, as done in clear~\cite{wu2019mantra}. Moreover, being able to identify the integrity threat from visually confidential image content should be relevant. Regarding the recognizability task, instead of only predicting the image class, we are planning to take an interest in object detection (its localization and its class) in protected images. Consequently, a subjective validation, involving human evaluation, should also be conducted.

\section*{Acknowledgment}
This work has been funded in part by the French National Research Agency (ANR-18-ASTR-0009), ALASKA project: https:// alaska.utt.fr, and by the French ANR DEFALS program (ANR-16- DEFA-0003).

\bibliographystyle{IEEEtran}
\bibliography{biblio}

\end{document}